\documentclass[graybox]{svmult}

\usepackage[caption=false]{subfig}
\usepackage{textcomp}
\usepackage[binary-units=true]{siunitx}
\usepackage[scaled=.92]{helvet} 
\usepackage{graphicx} 
\usepackage{balance}  
\usepackage{booktabs} 
\usepackage{ccicons}  
\usepackage{ragged2e} 
\usepackage{hyperref}
\usepackage[utf8]{inputenc}

\def\plaintitle{A Multi-Media Exchange Format for Time-Series Dataset Curation}
\def\plainauthor{Philipp M. Scholl, Benjamin Völker, Bernd Becker, Kristof Van Laerhoven}
\def\emptyauthor{}
\def\plainkeywords{Data Curation; Activity Recognition; Multi-Media Format; Data Storage; Comma-Separated-Values}
\def\plaingeneralterms{Documentation, Standardization}

\definecolor{linkColor}{RGB}{6,125,233}
\hypersetup{%
  pdftitle={\plaintitle},
  pdfauthor={\emptyauthor},
  pdfkeywords={\plainkeywords},
  bookmarksnumbered,
  pdfstartview={FitH},
  colorlinks,
  citecolor=black,
  filecolor=black,
  linkcolor=black,
  urlcolor=linkColor,
  breaklinks=true,
}

\begin{document}
\def\plaintitle{A Multi-Media Exchange Format for Time-Series Dataset Curation}
\def\plainauthor{Philipp M. Scholl, Benjamin Völker, Bernd Becker, Kristof Van Laerhoven}
\def\emptyauthor{}
\def\plainkeywords{Data Curation; Activity Recognition; Multi-Media Format; Data Storage; Comma-Separated-Values}
\def\plaingeneralterms{Documentation, Standardization}

\title*{\plaintitle}
\titlerunning{\plaintitle}
\author{\plainauthor}
\institute{Philipp M. Scholl,  Benjamin Völker and Bernd Becker \at University of Freiburg, \email{<pscholl,voelkerb>@informatik.uni-freiburg.de}
\and Kristof Van Laerhoven \at University of Siegen \email{kvl@eti.uni-siegen.de}}
%
%
\maketitle


\abstract{
  Exchanging data as character-separated values (CSV) is slow, cumbersome and error-prone.
  Especially for time-series data, which is common in Activity Recognition, synchronizing several independently recorded sensors is challenging.
  Adding second level evidence, like video recordings from multiple angles and time-coded annotations, further complicates the matter of curating such data.
  A possible alternative is to make use of standardized multi-media formats.
  Sensor data can be encoded in audio format, and time-coded information, like annotations, as subtitles.
  Video data can be added easily.
  All this media can be merged into a single container file, which makes the issue of synchronization explicit.
  The incurred performance overhead by this encoding is shown to be negligible and compression can be applied to optimize storage and transmission overhead.
}


\section{Introduction}
 At the heart of each Activity Recognition task is a dataset.
 This dataset might be formed from multiple media streams, like video, audio, motion and other sensor data.
 Recorded at different rates, sparsely or uniformly sampled and with different numerical ranges, these streams are challenging to process and store.
 Commonly, datasets are published in multiple character-separated values (CSV) files, either with a constant rate or time-coded.
 For small, independent time-series this is a worthwhile approach, mostly due to its simplicity and universality.
 However, when observing with multiple independent sensors, synchronization quickly becomes a challenge.
 Different rate recordings have to be resampled, time-coded files have to be merged.
 Storing such data in several (time-coded) CSV files hides this issue, until the dataset is going to be used.
 Furthermore parsing CSV files incurs a large performance and storage overhead, compared to a binary format.

  \vspace{\baselineskip}
  \fbox{%
    \begin{minipage}{0.925\linewidth}
      \textbf{Examples of Activity Recognition Datasets} \vspace{-.3pc} \\
      
      \textbf{HASC Challenge \cite{Kawaguchi2011}:} \textgreater 100 subjects, time-coded CSV files.  \\
      \textbf{Box/Place-Lab \cite{Intille2006}:} A sensor-rich home, in which subjects are monitored for long terms. Data is available in time-coded CSV files.   \\
      \textbf{Opportunity \cite{Roggen2010}:} 12 subjects were recorded with 72 on- and off-body sensors in an Activities of Daily Living (ADL) setting. Multiple video cameras were used for post-hoc annotations. Data is published in synchronized CSV files.  \\
      \textbf{Kasteren's Home \cite{Kasteren2010}:} 12 sensors in 3 houses. Data is stored in matlab files. \\
      \textbf{Borazio's Sleep \cite{Borazio2014}:} 1 sensor, 42 subjects. Data is stored in numpy's native format. \\
      \textbf{Freiburg Longitudinal \cite{kristof}:} 1 sensor, 1 subject, 4 weeks of continuous recording. Data is stored in numpy's native format. \\
    \end{minipage}}\label{sec:sidebar}
  \vspace{\baselineskip}

 One alternative is storing such datasets in databases, like SQL, NOSQL or HDF5-formats.
 This eases the management of large datasets, but shows the same issues as a CSV format, namely that there is no direct support for time-series or video data.
 An alternative approach is to store time-series in existing multi-media formats.
 Encoding all multi-media data in one file allows to merge streams, to synchronize them and to store (meta-)data in a standardized format.
 In the next section, we will first look at the formats commonly used to exchange data in Activity Recognition, afterwards we detail a multi-media format and evaluate the incurred performance and storage overhead of each format.

\section{Related Work}

 In the classic activity recognition pipeline \cite{Bulling2014}, the first step is to record and store sensor data. 
 The observed activities, executed by humans, animals or other actors, are recorded with different sensors.
 Each sensor generates a data \emph{stream}, whether this is a scene camera for annotation purposes, a body-worn motion capturing system or binary sensors like switches.
 Sampled at different rates, with differing resolutions, ranges, units and formats these streams offer a large variety of recording parameters.
 These parameters are usually documented in an additional file that resides next to the actual data 
 The actual data is commonly stored in a CSV file, in a binary format for Matlab or NumPy, or in format specific to some Machine Learning framework like ARFF \cite{Hall2009} or libSVM \cite{Chang2011}.

 Synchronizing such multi-modal data, i.e.~converting this data to the same rate and making sure that recorded events happened at the same time presents a major challenge 
 Possible approaches range from offline recording with post-hoc synchronization on a global clock, to live streaming with a minimum delay assumption - all but the last one require some form of clock synchronization and careful preparation.
 Storing events with timestamps on a global clock is then one possible way to allow for post-recording synchronization, i.e.~each event is stored as a tuple of \texttt{\textless timestamp, event data\textgreater}.

 The subsequent step of merging such time-coded streams often requires to adapt their respective rates.
 Imagine, for example, a concurrent recording of GPS at $3 \si{Hz}$ and acceleration at $100 \si{Hz}$.
 To merge both streams:
  will GPS be upsampled or acceleration downsampled, or both resampled to a common rate?
  Which strategy is used for this interpolation, is data simply repeated or can we assume some kind of dependency between samples?
  How is jitter and missing data handled?
  These question need to be answered whenever \emph{time-coded} sensor data is used.
  A file format which makes the choice of possible solutions explicit is the goal of this paper.

\section{Multi-Media Container Approach}

 Sensor data commonly used in Activity Recognition is not different from low-rate audio or video data.
 Common parameters are shared, and one-dimensional sensor data can be encoded with a lossless audio codec for compression.
 Rate, sample format and number of channels need to be specified for an audio track.
 The number of channels is equivalent to the number of axis an inertial sensor provides, as well as its sample rate.
 The sample format, i.e.~how many bits are used to encode one measurement, is also required for such a sensor.
 Other typical parameters, like the range settings or conversion factor to SI units (if not encoded as such), can be stored as additional meta-data, as those are usually not required for an audio track.

 Lossless compression, like FLAC \cite{flac} or WavPack \cite{wavpack}, can be applied to such encoded data streams.
 This allows to trade additional processing for efficient storage.
 In the evaluation section several lossless schemes are compared.
 These include the general LZMA2 and ZIP compressors, and the FLAC \cite{flac} and WavPack \cite{wavpack} audio compressors.
 All but the first two can be easily included in multi-media container formats.
To use audio streams, data needs to be sampled at a constant rate, i.e. the time between two consecutive samples is constant and only jitter smaller than this span is allowed.
 Put differently, the time between two consecutive data samples $t_{i}$ and $t_{i+1}$ at frame $i$ must always be less than or equal to the sampling rate $r$: $\forall i \in N: t_{i+1} - t_{i} \leq \frac{1}{r}$.
 Compared to time-coded storage, the recording system has be designed to satisfy this constraint.
 Problems with a falsely assumed constant rate recording setup will therefore surface faster.
 Especially in distributed recording settings, where the just mentioned constraint is checked only against local clocks which might drift away from a global clock, is a common challenge.

 Sparsely sampled events can be encoded as subtitles.
 Here, each sample is recorded independently of its preceding event, i.e.~the above mentioned constraint does not hold.
 Each event needs to be stored with a time-code and the actual event data.
 Depending on the chosen format, this can also include a position in the frame of an adjacent video stream or other information.
 For example, this can be used to annotate objects in a video stream.
 A popular format is the Substation Alpha Subtitle (SSA\cite{substation}) encoding, which includes the just mentioned features.
 Since data is encoded as strings, it is suitable for encoding ground truth labels.
 To a limited extent, since no compression is available, it can be used for sensor events as well.
 For example, low rate binary sensors, like RFID readers could be encoded as subtitles.

 Encoded sensor and subtitle data can then be combined with audio and video streams in a multi-media container format.
 One such standard is the Matroska \cite{matroska} format, that is also available in a downgraded version called WebM \cite{webm} for webbrowsers.
 Once the data streams are combined into one such file, this data can be "played" back in a synchronous manner.
 This means that streams recorded at different rates, and in different formats, need to be converted to a common rate and possibly common format.
 Meta-data that contains additional information like recording settings, descriptions and identifiers can be stored in addition to the parameters already contained in the stream encoding.
 For this task off-the-shelf software, like FFmpeg \cite{ffmpeg} can be used, which also provides functionality like compression, resampling, format conversion and filtering.
 Annotation tasks can be executed with standard subtitle editing software, discouraging the creation of yet another annotation tool.
 Furthermore, video streaming servers can be used for transporting live sensor data recordings to remote places.

 The use of such a standard format for curating datasets allows for re-using existing software, however not without limitations.
 Asynchronous, also called sparsely sampled, data recorded at high rates is not supported.
 This mainly stems from the simplifying assumption that streams are recorded with a constant rate.
 We presume that satisfying this constraint while recording to be easier than handling asynchronicity later on.
 For example, breaks, shifts or jitter due to firmware bugs can be detected earlier.
 In general this is a hard limitation, however different data types can also be encoded in multiple streams.
 Also, the en- and decoding overhead might be a limitation, which we will look at in the next section.

\begin{figure}
  \centering
  \subfloat[storage efficiency] {\centering
    \includegraphics[width=.49\linewidth]{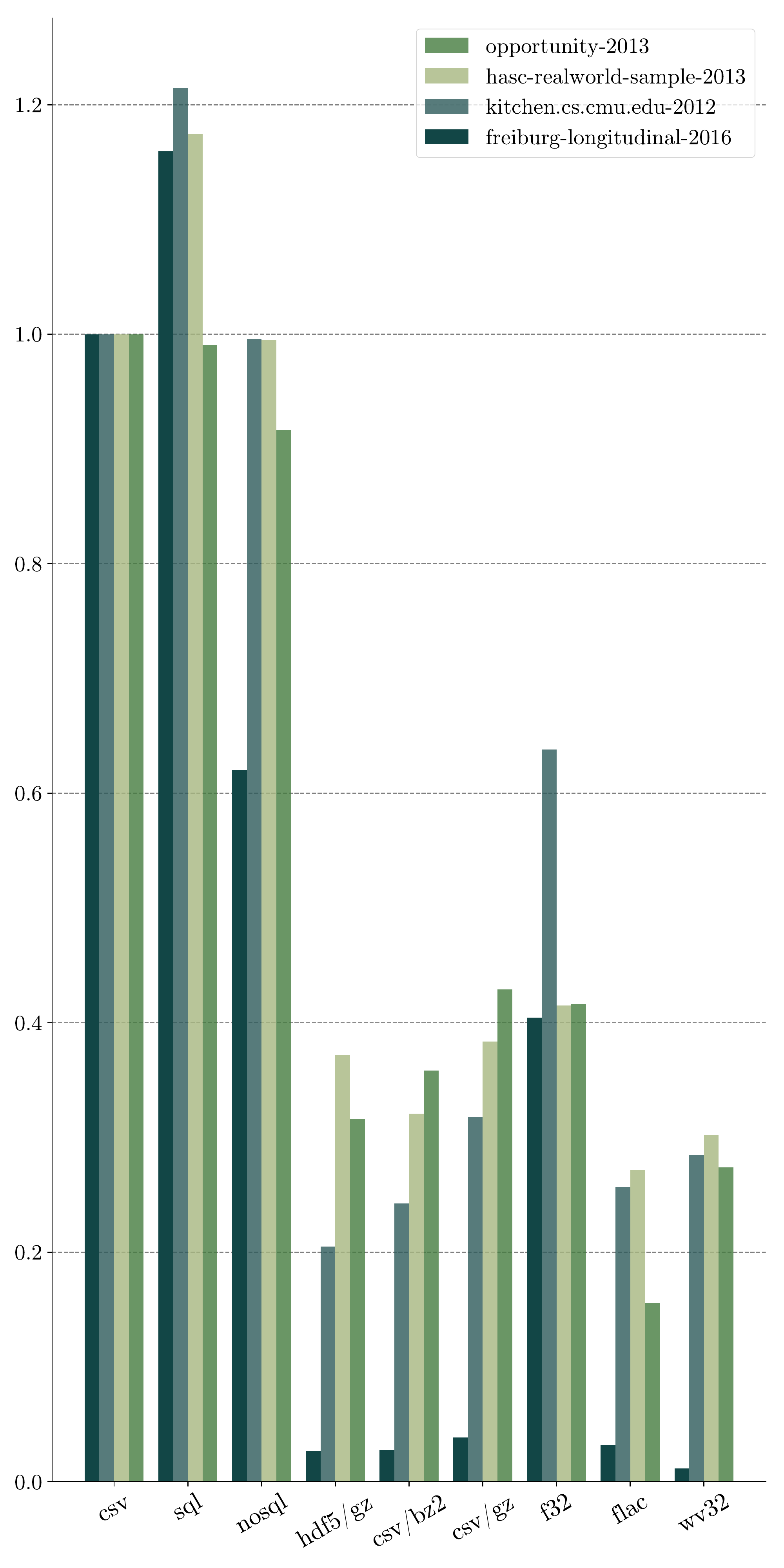}
    \label{fig:compression} }
  \subfloat[runtime overhead]{
    \centering
    \includegraphics[width=.49\linewidth]{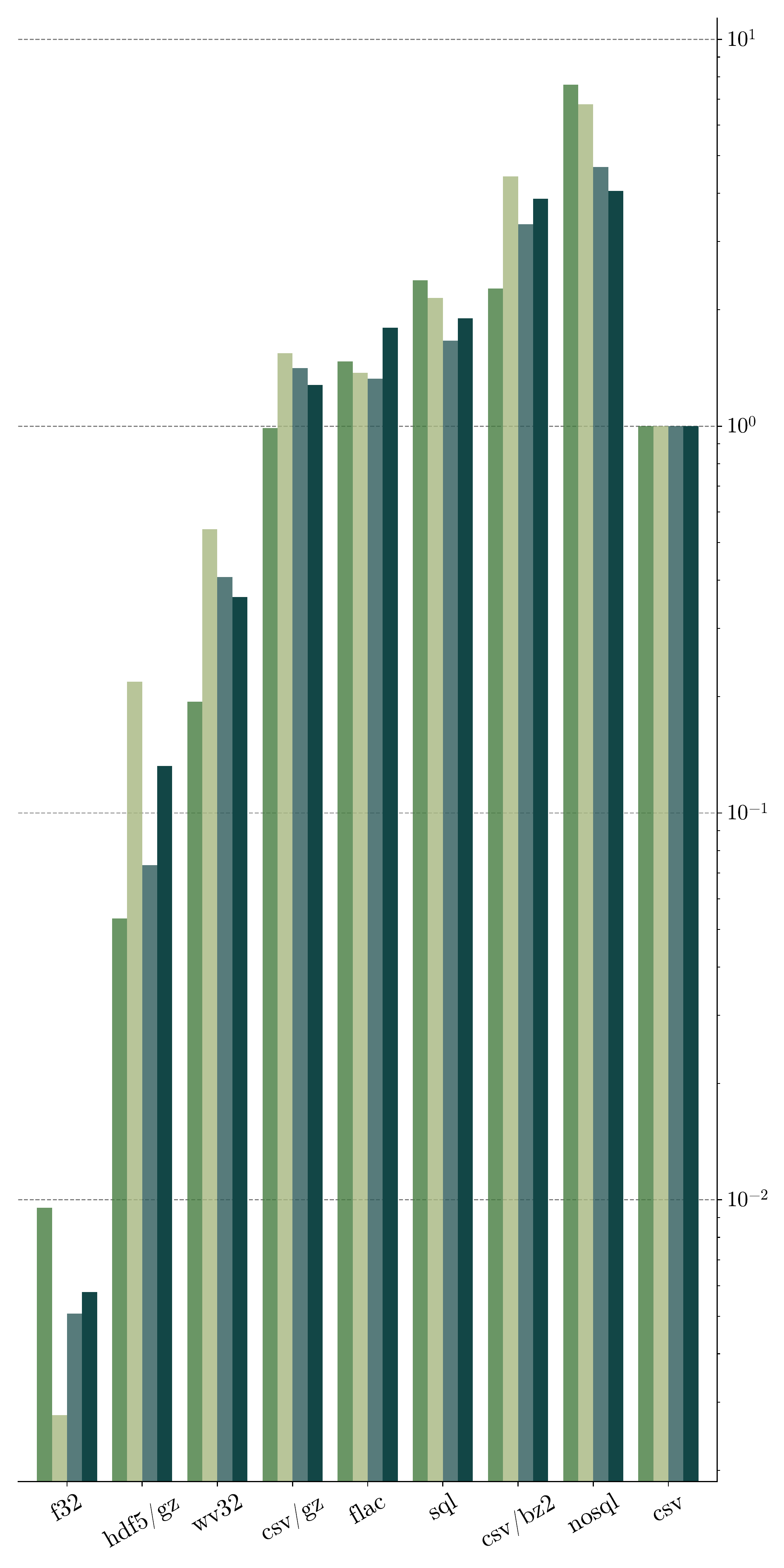}
    \label{fig:overhead}
  }
  \caption{Comparison of runtime overhead and storage efficiency, which is to be traded off for each format. Each is given relative to a deflated CSV file.}
\end{figure}

 Compared to the de-facto standard of using CSV files, encoding sensor data as audio, annotations as subtitles and audio- and video-data in standard formats provides several improvements.
 Important parameters like sampling rate, format and number of axes are included in the file.
 Adding additional information as meta-data leads to a compleltly \emph{self-descriptive} format.
 \emph{Synchronous} playback of multiple streams, which requires re-sampling, is supported by off-the-shelf software.
 Related problems, like un-synchronized streams can be caught earlier, since this step is explicit.
 The container format is \emph{flexible} enough to support different number formats, i.e.~values can be encoded as floats or integers of varying bit-size.
 Optional compression leads to \emph{compact} storage, which allows for efficient storage and transmission.
 Additionally, when thinking about large datasets, such a container format requires \emph{divisible} storage.
 This functionality (seeking without reading the whole dataset into memory\footnote{which would be required for time-coded storage}) is provided. Such divisible storage also allows for streaming applications, which, for multi-media format, also provides network protocols to cover transmission via un-reliable links.

\section{Evaluation}

 Compressing sensor data as an audio stream incurs an en- and decoding overhead, and provides optimized storage.
 In this section both are quantified.
 By a repetitive measurement of the relative wall clock time for decompression, the processing overhead is measured.
 This runtime overhead is reported as the fraction of reading time relative to reading and \emph{converting} the CSV file into main memory.
 The compression factor is determined by comparing the number of bytes required to store the compressed file to the original, deflated CSV file.
 Binary and text-based storage is compared.
 The Zip and LZMA2 algorithms are used for general byte-wise compression, and the lossless FLAC and WavPack compressor for audio-based compression.
 Additionally storing in the sqlite3, and MongoDB database, as well as the HDF5 format is compared.
 The approach of compressing binary files with a general compressor is used by Numpy, MongoDB and HDF5 for example.
 
 The test were run on the Kitchen CMU \cite{de2009guide}, Opportunity \cite{Roggen2010}, HASC Challenge \cite{Kawaguchi2011} and on twenty days of the Freiburg Longitudinal Wrist \cite{kristof} datasets.
 A machine with an i7-4600U CPU running at $2.1 \si{\giga \hertz}$ with $8 \si{\giga \byte}$ of memory was used for all tests.
 \autoref{fig:overhead} and \autoref{fig:compression} show the results of these tests, \emph{csv/gz} refers to a zip-compressed CSV file, \emph{csv/bz2} to an LZMA2 compressed file\footnote{the XZ utils package was used}, \emph{f32} refers to a $32-\si{\bit}$-float binary format, \emph{wv32} to WavPack compression of $32-\si{\bit}$-floats, and \emph{flac} to the \emph{FLAC} compressor which only supports $24\si{bits}$ values. \emph{nosql} refers to storing the data in a MongoDB (\emph{NoSQL}) database, \emph{sql} to sqlite3 storage, and \emph{hdf5/gz} to encoding the sensor data in a zip-compressed HDF5 container.
 For \emph{MongoDB} storage, each data stream is stored together with its metadata into a so called \emph{collection} inside the database. Since MongoDB's \emph{BSON} format only allows for certain data types, we choose the $64\si{bits}$ \emph{Double} format to store the individual event data with the default compression parameters. 
 Equivalent to the \emph{MongoDB} structure, each datastream is stored as a separate table in the \emph{sqlite} database and the event data is stored in $64\si{bits}$ \emph{REAL} format. The stream's metadata is stored in a separate table. 
 De- and encoding was performed using the corresponding python interfaces \emph{pymongo} and  \emph{sqlite3}. 
 The \emph{HDF5} \cite{hdf5} data was generated and de-/encoded using the \emph{h5py} python interface and stored with zip-compression.

\subsection{Processing Overhead}

 It is interesting to check how much overhead is incurred for decompression by each storage paradigm, as this gives an insight if data needs to be stored in an intermediate format while evaluating a recognition pipeline.
 If the overhead is comparatively low, no intermediate storage format is required and data can always be loaded from such a file.
 However, should decoding require more processing time than the actual processing, an intermediate file needs to be used.

 Naturally such a format would be binary, at best a memory image which can be mapped into main memory as a $32 \si{\bit}$-float.
 The baseline is, therefore, the time required to convert a CSV from disk into a binary format in memory.
 The fraction of time required to do the same for each storage scheme is reported in \autoref{fig:overhead}.
 Each test is repeated six times, and the first run is discarded, i.e.~data is always read from the disk cache.

 Just parsing a CSV file incurs an up to hundred-fold overhead compared to reading a binary file (\emph{f32} in \autoref{fig:overhead}).
 Compressing CSV data\footnote{note that the \autoref{fig:overhead} represent the factor between the compression and simply \emph{reading} an uncompressed CSV file} can increase the runtime by $1.4 - 3.0$ times.
 So, looking only at runtime performance a CSV or compressed CSV file should hardly be used for large datasets.
 When comparing compression schemes, it can be seen that a $32 \si{\bit}$ WavPack compression provides the second lowest runtime overhead, only the hdf5-scheme is roughly twice as fast.
 Storing data into a \emph{MongoDB} database comes with the cost that interfacing is done over a TCP socket in plaintext JSON format, which incurs an up to $4$-times overhead compared to CSV, and at least $100$-fold compared to raw binary storage. 
 A trade-off between storage efficiency and performance has to be found.

\subsection{Storage Efficiency}
 General compression and audio compression algorithms were tested.
 Raw binary, WavPack \cite{wavpack} and FLAC \cite{flac} compression were taken from the FFmpeg \cite{ffmpeg} suite with default parameters.
 \autoref{fig:compression} shows the amount of compression that was achieved for each dataset per algorithm compared to uncompressed CSV files.

 The used datasets show different characteristics found in other datasets as well.
 For example the Longitudinal \cite{kristof} dataset can be massively compressed with general algorithms, almost down to $2\%$ of its original size.
 This is mainly owed to the fact that the contained acceleration data was recorded with a resolution of only 8-bits, and that a run-length compression was already applied during recording.
 This run-length compression is deflated for CSV storage first, adding a lot of redundancy.
 Generally, text formats provide a larger storage efficiency only when less characters are required per sample than their binary counterparts.

 This effect is partially visible for the kitchen dataset \cite{de2009guide}.
 The relative storage requirements for binary storage (f32 in \autoref{fig:compression}) is a lot larger than for other datasets.
 Here, the text representation of values is smaller since they range from $0-1$, i.e. the text values are almost always of the same length.
 Other datasets provide a range of values with larger pre-decimals, hence a longer text representation.
 The maximum dynamic range that can be stored with a text-based format more efficiently is therefore limited to the (decimal) encoding, (less than $10000$ for five digits), while a comparable binary encoding can range up to $2^{5*8}$.

 When optimizing data for space efficiency, the encoding of each value is the most critical factor.
 Limiting the number of \si{bits} per value, in essence assuming a limited dynamic range of the encoded signal, has the strongest influence on the storage efficiency. 
 The 24 bit flac encoding shows the best overall efficiency due to this.
 If a dynamic range of more than $2^{24}$ is required, the wavpack encoding should be considered. 
 However, when encoding values in text format with a limited dynamic range (equivalent to four characters), a text compression algorithm is not worse than encoding data in binary format.
 For the general case and when binary storage can be used, the WavPack compression provides the same storage efficiency as the more general LZMA2 compressor and provides comparable compression factors to the hdf5 format. 
 
 NoSql, and SQL storage do hardly provide a benefit over CSV concerning storage efficiency in our tests.
 For SQL, this is probably an implementation artifact, since sqlite3 stores date internally as a string representation.
 NoSql, represented via MongoDB here, stores data in a compressed binary JSON format.
 This provides a benefit only for one dataset, which contains a lot of redundancy.

\section{Conclusion}

 Curated time-series data provides the basis for comparing Activity Recognition and other Machine Learning approaches in an objective and repeatable manner.
 This data usually includes low-rate sensor data, e.g.~motion sensors, time-coded annotations, and second-level evidence like video and audio recordings.
 The state of the art for exchanging this data seems to be a mix of time-coded CSV format, and dedicated audio- and video-codecs.
 Synchronizing the stored data-streams is usually not done by the dataset provider and the dataset consumer is left with this challenge that usually requires information of the recording setup.
 This is especially problematic when video or audio data is recorded in addition to sensor data.
 The HDF5 which can provide a smaller decoding overhead, provides no support for decoding audio- and video-data directly.

 The CSV format incurs a large overhead both in runtime and storage.
 A possible alternative, with lower overhead, is presented here.
 Motion and other sensor data, as well as extracted features, can be stored in lossless audio formats.
 Ground truth labels and other time-coded information can be stored in subtitle format.
 These streams can then be merged in a common multi-media container (e.g.~Matroska), with additional video streams.
 One recording session is stored in a single file, that can be \emph{self-descriptive}, \emph{synchronized}, with a fitting \emph{storage-runtime trade-off} for multi-media data and supports \emph{live-streaming}.




\bibliographystyle{splncs04}
\bibliography{hasca2019}
\end{document}